\magnification=1200
\baselineskip=18pt
\input epsf
\def\preprint{Y}

\def\im{{\rm\ Im\ }}
\def\tr{{\rm\ Tr\ }}
\def\dirac{{\bf\rm D}}
\def\wilson{{\bf\rm B}}
\def\ham{{\bf\rm H}}
\def\mbham{{\cal H}}
\def\wblUp{{{}^{\rm WB}_{\ \ U^\prime}\!\langle}}
\def\wbrUp{{\rangle^{\rm WB}_{U^\prime}}}
\def\wblU{{{}^{\rm WB}_{\ \ U}\!\langle}}
\def\wbrU{{\rangle^{\rm WB}_U}}
\def\wbl1{{{}^{\rm WB}_{\ \ 1}\!\langle}}
\def\wbr1{{\rangle^{\rm WB}_1}}
\def\cap{\hsize=4.5in}

\def\figure#1#2#3{\if \preprint Y \midinsert \epsfxsize=#3truein
\centerline{\epsffile{figure_#1_eps}} \halign{##\hfill\quad
&\vtop{\parindent=0pt \hsize=5.5in \strut## \strut}\cr {\bf Figure
#1
}&#2 \cr} \endinsert \fi}

\def\captionone{\cap
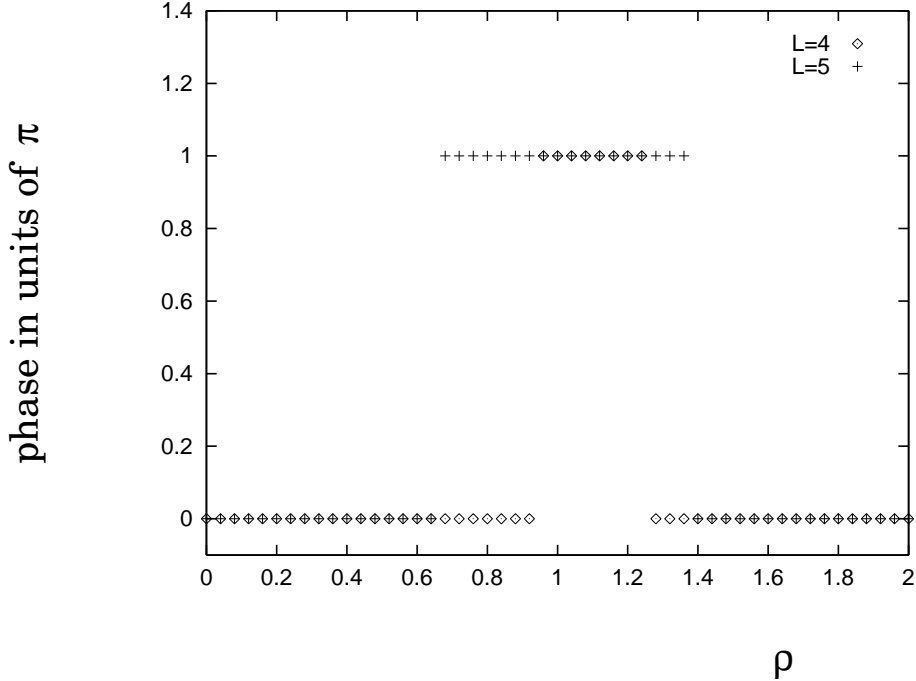
Plot of the phase of the overlap on the trivial orbit as a function
of a gauge transformation $g_n$ parameterized by $\rho$.
}
\def\captiontwo{\cap
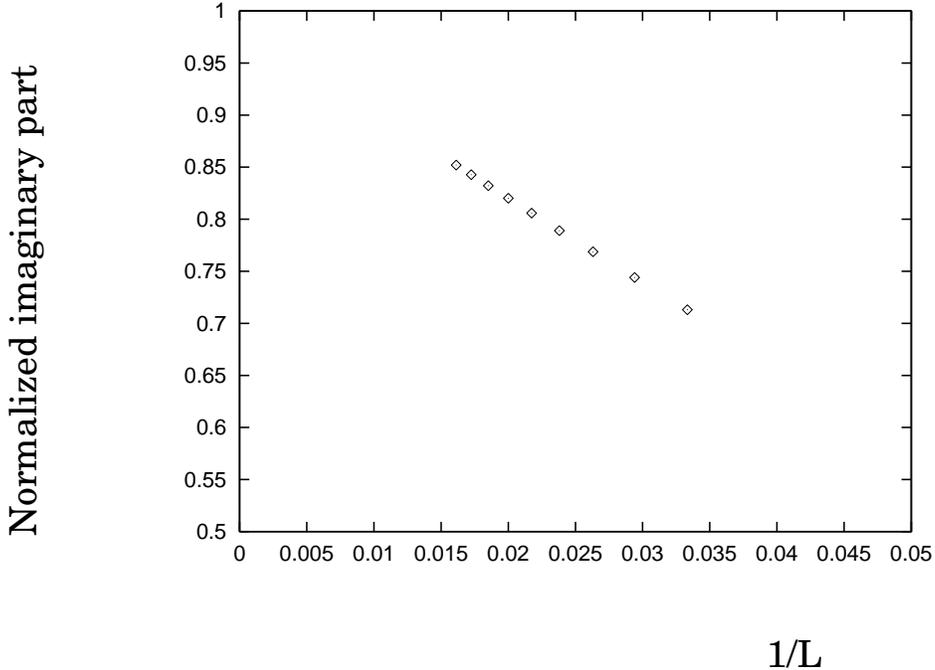
Plot of the normalized imaginary part of the
action as a function of ${1\over L}$ 
at $m_p=6$.
}
\def\captionthree{\cap
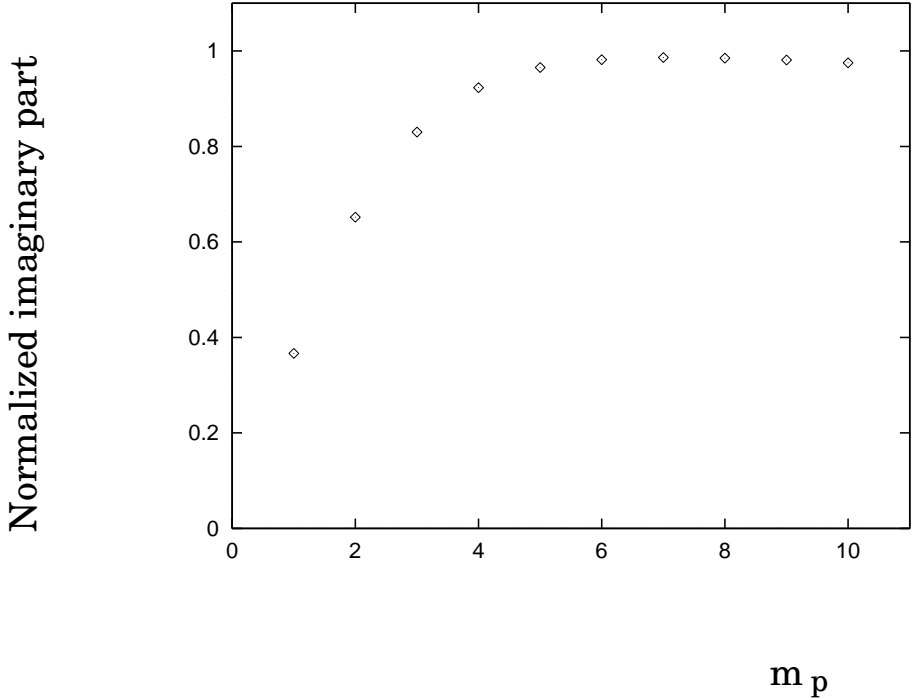
Plot of the normalized imaginary part of the
action at $L=\infty$ as a function of $m_p$.
}
\rightline{DOE/ER/40561-320-INT97-00-168}
\rightline{UW/PT-97-06}
\rightline{DPNU-97-19}
\rightline{August 1997}
\medskip
\centerline{\bf Parity invariant lattice regularization of}
\centerline{\bf three-dimensional gauge-fermion system}
\vskip 0.5truecm
\centerline{Rajamani Narayanan}
\centerline {\sl Institute for Nuclear Theory, Box 351550}
\centerline {\sl University of Washington, Seattle, WA 98195-1550, USA}
\centerline{e-mail: {\tt rnarayan@fermi.phys.washington.edu}}
\vskip 0.5truecm
\centerline{Jun Nishimura}
\centerline {\sl Department of Physics, Nagoya University}
\centerline {\sl Chikusa-ku, Nagoya 464-01, Japan}
\centerline{e-mail: {\tt nisimura@eken.phys.nagoya-u.ac.jp}}
\vskip 1truecm
\vfill
\centerline{\bf Abstract}
\vskip 0.5truecm
In three dimensions,
the effective action for the gauge field induced by integrating out 
a massless Dirac fermion
is known to give either a parity-invariant or a parity-violating result, 
depending on the regularization scheme. 
We construct a lattice formulation of the massless Dirac fermion
using the overlap formalism. We show that the result is parity
invariant in contrast to the formulation using Wilson fermions
in the massless limit. This facilitates a non-perturbative
study of three-dimensional massless Dirac
fermions interacting with a gauge field in a
parity invariant setting with no need for fine-tuning.
\bigskip
\leftline{PACS: 11.15.Ha; 11.10.Kk; 11.30.Er}
\leftline{Keywords: Lattice gauge theory, three dimensions, parity}
\medskip

\vfill\eject

\bigskip
\centerline{\bf I. Introduction}
\medskip
Three-dimensional gauge-fermion system has been studied intensively
in the context of fractional quantum Hall effect [1]
and high $T_c$ superconductivity [2].
The gauge boson can acquire a mass 
without violating the gauge symmetry due to the existence of 
the Chern-Simons term [3].
The Chern-Simons term affects the statistics of the matter field,
giving rise to fermion-boson transmutation
and exotic excitations known as anyons.
A peculiar phenomenon such as the breakdown of 
Lorentz invariance through spontaneous magnetization is also conjectured 
[4].

It is well known that
the effective action for the gauge field induced by integrating out
the massless Dirac fermion can either be invariant under 
parity\footnote{${}^\star$}
{In three dimensions, parity refers to the reflection about all three
space coordinates.} 
or break the parity symmetry, depending on the regularization procedure.
If parity is a symmetry of the regularization then it cannot preserve
gauge invariance under certain large gauge transformations allowed
by the gauge group. 
It is conventional to refer to the breaking of the parity 
symmetry or the gauge invariance in
the massless limit as ``parity anomaly'' [5] although it is a
regularization dependent statement.
Decoupling of the fermion in the infinite mass theory is also regularization
dependent. 

The fermion integral for a massless Dirac fermion
can be regularized by the introduction of a
single Pauli-Villars field. The mass of the Pauli-Villars field
breaks the parity and its effect remains even when one takes the
mass to infinity thereby yielding a parity-violating result for
the integral over fermions.
The coefficient of the parity odd term in the effective action
changes its sign according to the sign of the Pauli-Villars mass,
and this is a sign of the regularization dependence. By adding more
Pauli-Villars field with appropriate signs for the mass term it
is possible to change the magnitude of the coefficient of the
parity odd term. 
One can adopt dimensional regularization instead,
regarding the totally anti-symmetric epsilon tensor as a purely 
three-dimensional object following the prescription proposed by
't Hooft-Veltman [6] and systematized by Breitenlohner-Maison [7].
The result comes out to be parity invariant.
Some other regularizations are also
known to give parity invariant result [8]. 

In all the above regularization schemes one can also study the case
when the Dirac fermion is massive. It is natural to expect the
fermion to completely decouple from the theory when its mass goes
to infinity. But this is not the case. In particular if the 
regulator is parity invariant in the massless limit it violates parity
in the limit of infinite mass indicating that the fermion has not
completely decoupled from the theory. In the case of Pauli-Villars
regulator it is possible to decouple the fermion in the limit of
infinite mass by a proper choice of the Pauli-Villars mass but 
the resulting theory will violate parity in the massless limit.

Lattice formalism is desired when one attempts a nonperturbative study
of the whole system including the gauge dynamics as well.
The standard Wilson's formalism for dealing with fermions on the lattice
has been examined by Coste-L\"uscher [9] and it has been shown 
that the result turns out to be parity violating in 
the massless limit.
The coefficient of the parity violating term changes its sign
according to the sign of the Wilson term and this is similar
to the result depending on the sign of the Pauli-Villars mass.
This can be easily understood by considering the signs of the masses
that doublers acquire through the Wilson term.
It would be interesting to obtain a lattice formalism that is
parity invariant for a massless Dirac fermion.
Such a formalism would result in a fermionic determinant that is real
but can change its sign under a specific large gauge transformation.
In this scheme, one can investigate if
the fermion decouples in the infinite mass limit or not.
This is the aim of the present paper.

The paper is organized as follows.
In Section II, we sketch the perturbative computation
of the effective action 
for the gauge field induced after the integration 
of the Dirac fermion,
using a single Pauli-Villars regulator
as well as dimensional regularization. We discuss the massless fermion
and also the infinitely massive fermion.
In Section III, we show that even within the Pauli-Villars regularization,
one can construct a parity invariant formulation for a massless fermion,
if one uses infinitely many Pauli-Villars fields. 
The need for infinite number of regulator fields is essential.
The situation is quite similar to the one encountered
when one regularizes a chiral fermion in even dimensions
using infinitely many Pauli-Villars fields [10,11].
In Section IV, we construct 
the overlap formalism [12]
for a three dimensional
massless Dirac fermion coupled to an external gauge field on the lattice.
We prove that the fermion determinant in the overlap formalism
is parity invariant.
In Section V, we describe the massive fermion within 
the overlap formalism to investigate the limit of infinite mass.
We calculate the effective potential under a constant 
background gauge field and see that the result in the infinite mass limit
is non-zero and parity odd. The parity odd term
is consistent with the one obtained in dimensional regularization,
which does not break parity in the massless limit.
Section VI is devoted to summary and conclusions.

\bigskip
\centerline{\bf II. Regularization dependence of the imaginary part
of the induced action}
\medskip

In the following, we consider the gauge group to be SU($N$) and
the fermion to be Dirac spinor in the fundamental representation.

The Euclidean action for a massive Dirac fermion
in the fundamental representation coupled to a SU($N$) gauge field 
is
$$S[\psi,\bar{\psi},A]
= \int d^3x \bar\psi(x) (\dirac+m)  \psi(x),  \eqno{(2.1)}$$
$$\dirac = \sum_{\mu=1}^3 \gamma_\mu (\partial_\mu + i A_\mu(x)).
\eqno{(2.2)}$$
In our notation, $A_\mu(x)$ is a finite
hermitian $N\times N$ matrix for each $x$ and $\mu$ and
$$\gamma_1= \pmatrix{0 & 1 \cr 1 & 0 \cr};\ \ \ \
\gamma_2= \pmatrix{0 & -i \cr i & 0 \cr};\ \ \ \
\gamma_3= \pmatrix{1 & 0 \cr 0 & -1 \cr}.\eqno{(2.3)}$$
The effective action $\Gamma(A)$ for the gauge field can be defined
formally as
$$
\eqalign{ e^{-\Gamma(A)}=
&
{ \int {\cal D} \psi {\cal D} \bar{\psi}
e^{-S[\psi,\bar{\psi},A]}
\over
{ \int {\cal D} \psi {\cal D} \bar{\psi}
e^{-S[\psi,\bar{\psi},0]} } }
\cr
=&
{ Det(D+m)\over Det(D_0+m)},
}\eqno{(2.4)} 
$$
where $D_0=D|_{A_\mu=0}$.
The behavior of the effective action $\Gamma(A)$ under parity transformation
of the gauge field:
$A_\mu(x) \rightarrow A_\mu^P(x):= -A_\mu(-x)$ is formally,
$$\Gamma(A^P)=\Gamma(A)^\ast,\eqno{(2.5)}$$
and this is satisfied by any
reasonable regularization of the effective action.
Therefore, the parity odd terms reside only in the imaginary part 
of the effective action.

When the fermion is massless, {\it i.e.,} $m=0$,
the action (2.1) is invariant under the parity transformation:
$$\psi(x) \rightarrow \psi(-x);\ \ \ \ \
\bar{\psi}(x) \rightarrow -\bar{\psi}(-x);\ \ \ \ \
A_\mu(x) \rightarrow -A_\mu(-x).\eqno{(2.6)}$$
One naively expects that the effective action is invariant
under the parity transformation of the gauge field:
$\Gamma(A^P)=\Gamma(A)$, which, combined with (2.5),
means that the effective action is real.
However, this is not necessarily true when the effective action is
properly regularized,
since the parity invariance (2.6) can be broken upon regularization.
When one cannot recover the parity invariance after removing the 
regularization with appropriate gauge-invariant counterterms, 
the theory is stated to have parity anomaly, which is signaled by
a non-vanishing imaginary part of the effective action.

The second order contribution to $\Gamma(A)$ can be written as
$$
\Gamma_2(A)=
{1\over 2} \int {d^3p  \over (2\pi)^3}
\Pi_{\mu\nu}(p) 
\tr [\tilde{A}_\mu(-p) \tilde{A}_\nu(p)],
\eqno{(2.7)}$$
where the trace is over the gauge group index.
The Fourier transform is defined through
$$
A_\mu(x) = 
\int {d^3q \over (2\pi)^3}
e^{iqx} \tilde{A}_\mu(q).
$$
The $\Pi_{\mu\nu}(p)$, which is called as vacuum polarization tensor,
can be given as follows.
$$ 
\Pi_{\mu\nu}(p) =
 {1\over 8}
\int {d^3q \over (2\pi)^3}
{\tr \bigl[
\gamma_\mu (-i\gamma_\lambda ({p-q\over 2})_\lambda + m)
\gamma_\nu
(i\gamma_\rho ({p+q\over 2})_\rho + m ) \bigr]
\over
\bigl[({p-q\over 2})^2 + m^2 \bigr]
\bigl[({p+q\over 2})^2 + m^2 \bigr]},
\eqno{(2.8)}$$
where the trace here is over the spinor index.
{}From rotational invariance,
$\Pi_{\mu\nu}(p)$ can be written as follows.
$$
\Pi_{\mu\nu}(p)
=
p_\mu p_\nu A(p^2) + \delta_{\mu\nu} B(p^2)
+  \epsilon_{\mu\nu\lambda} p_\lambda C(p^2).
\eqno{(2.9)}
$$
{}From power counting, $B(p^2)$ is linearly divergent, and hence
necessitates a regularization.
Note that the parity odd term $C(p^2)$ is already convergent on dimensional
grounds, since it contains a factor of $m$.

The effective action can be regularized using a 
single Pauli-Villars field $\phi(x)$, which has the 
following action:
$$S^{(b)} = \int d^3x \bar{\phi}(x) (\dirac+\Lambda)  \phi(x),  
\eqno{(2.10)}$$
where
$\Lambda$ is the mass of the Pauli-Villars field and $\bar{\phi}$ and $\phi$
are commuting variables.
The $\Pi_{\mu\nu}(p)$ is regularized as
$$\eqalign{\Pi_{\mu\nu}^{(reg)}(p)=&
 {1\over 8} 
\int {d^3q \over (2\pi)^3}
\Biggl[
{\tr \bigl[
\gamma_\mu (-i\gamma_\lambda ({p-q\over 2})_\lambda + m)
\gamma_\nu
(i\gamma_\rho ({p+q\over 2})_\rho + m ) \bigr]\over
\bigl[({p-q\over 2})^2 + m^2 \bigr]
\bigl[({p+q\over 2})^2 + m^2 \bigr]}
\Biggr.
\cr
~& -
\Biggl.
{\tr \bigl[
\gamma_\mu (-i\gamma_\lambda ({p-q\over 2})_\lambda + \Lambda )
\gamma_\nu
(i\gamma_\rho ({p+q\over 2})_\rho + \Lambda ) \bigr]\over
\bigl[({p-q\over 2})^2 + \Lambda^2 \bigr]
\bigl[({p+q\over 2})^2 + \Lambda^2 \bigr]}
\Biggr]  .
\cr
}\eqno{(2.11)}$$
Since this regularization preserves gauge invariance, we have
$$
\Pi_{\mu\nu}^{(reg)}(p)
=
(p_\mu p_\nu - p^2 \delta_{\mu\nu})
A^{(reg)}(p^2)
+  \epsilon_{\mu\nu\lambda} p_\lambda C^{(reg)}(p^2).
\eqno{(2.12)}
$$
Hence the above expression is convergent
in the $\Lambda\rightarrow \infty$ limit.
The parity odd term is given by
$$
C^{(reg)}(p^2)
=
{m\over 2\pi \sqrt{p^2}} sin^{-1}\sqrt{p^2\over p^2 + 4 m^2}
-{ \Lambda\over 2 \pi \sqrt{p^2}} sin^{-1}\sqrt{p^2\over p^2 + 4 \Lambda^2}.
\eqno{(2.13)}
$$
The first term vanishes in the $m\rightarrow 0$ limit.
The second term gives $\mp {1 \over 4 \pi}$ 
in the $\Lambda \rightarrow \pm \infty$
limit.
Therefore the imaginary part of $\Gamma_2(A)$ is given by
$$
\im \Gamma_2(A) = \pm 
{1 \over 8 \pi } \int d^3x \epsilon_{\mu\nu\lambda} 
\tr (A_\mu(x) \partial_\nu A_\lambda(x)).
\eqno{(2.14)}
$$
Thus we have parity-violating result.

On the other hand,
$ \lim_{(m,\Lambda)\rightarrow \infty} C^{(reg)}(p^2) = 0$
indicating that there is no parity breaking term in the infinite mass
limit, if we take the Pauli-Villars mass to have the same sign as the fermion
mass. Further, we have
$$ \lim_{m\rightarrow\infty,\Lambda\rightarrow\pm\infty} C^{(reg)}(p^2) -  
\lim_{m\rightarrow 0,\Lambda\rightarrow\pm\infty} C^{(reg)}(p^2) 
= 
{1 \over 8 \pi } \int d^3x \epsilon_{\mu\nu\lambda} 
\tr (A_\mu(x) \partial_\nu A_\lambda(x)).
\eqno{(2.15)}$$
and this result will be regularization independent.

The theory can also be regulated using dimensional regularization.
Here the existence of the 
parity-violating object $\epsilon_{\mu\nu\lambda}$
makes the regularization nontrivial.
We follow the prescription proposed by 't Hooft-Veltman [6]
and systematized by Breitenlohner-Maison [7].
We regard the external momentum $p_\mu$ to be three-dimensional
and the internal momentum $q_\mu$ to be $(3+\epsilon)$-dimensional.
We decompose $q_\mu$ as
$$
q_\mu = \tilde{q}_\mu + \hat{q}_\mu,
\eqno{(2.16)}
$$
where  $\tilde{q}_\mu$ and $\hat{q}_\mu$ are the projection 
onto the subspaces ${\cal R}^3$ and ${\cal R}^\epsilon$.
The formulae for the gamma matrices are 
$$
\eqalign{
\tr(\gamma_\mu)&=0 \cr
\tr(\gamma_\mu \gamma_\nu) & =\delta_{\mu\nu} \tr I  \cr
\tr(\gamma_\mu \gamma_\nu \gamma_\lambda) & = i \epsilon_{\mu\nu\lambda} 
\tr I \cr
\tr(\gamma_\mu \gamma_\nu \gamma_\lambda \gamma_\rho)
&= (\delta_{\mu\nu} \delta_{\lambda\rho}-
\delta_{\mu\lambda} \delta_{\nu\rho}+
\delta_{\mu\rho} \delta_{\nu\lambda})  \tr I .
\cr}
\eqno{(2.17)}
$$
We define $\tr I=2$.
The $\epsilon_{\mu\nu\lambda}$ is considered to be purely three dimensional:
$$
\epsilon_{\mu\nu\lambda} \hat{q}_\lambda = 0.
\eqno{(2.18)}
$$
As is obvious from above, 
this prescription does not preserve SO(3+$\epsilon$) invariance, but
only SO(3)$\times$SO($\epsilon$).
However this is the only algebraically consistent one
known so far, and it has been used successfully in some applications
[13].

Returning to our case, since this is also a gauge invariant regularization,
the decomposition (2.12) is valid, and the 
dimensionally regularized vacuum polarization tensor 
$\Pi^{(reg)}_{\mu\nu}(p)$ is convergent 
in the $\epsilon \rightarrow 0$ limit.
After taking the limit, we have the following for 
the parity violating part $C^{(reg)}(p^2)$.
$$
C^{(reg)}(p^2)
=
{m\over 2 \pi \sqrt{p^2}} sin^{-1}\sqrt{p^2\over p^2 + 4 m^2},
\eqno{(2.19)}
$$
which vanishes as $m\rightarrow 0$.
Thus the result is parity invariant.
It is clear that (2.15) remains the same in this regularization also.
Since the massless theory is parity invariant in this regularization
scheme, gauge invariance has to be broken but this cannot be seen in
perturbation theory where gauge invariance will be preserved.

To get the full expression for the imaginary part one also needs to
compute the cubic term in the gauge field. This can be deduced
by gauge invariance under small gauge transformations.
The results for the imaginary part of the effective action 
in the two limits are
$$\eqalign{
\lim_{m\rightarrow 0} \im \Gamma(A) & = 
\cases{ \pi n  S_{CS}(A) + h(A)  & if $n$ is even \cr
\pi n  S_{CS}(A) + h(A) & if $n$ is odd \cr}
\cr
\lim_{m\rightarrow \infty} \im \Gamma(A) & =
\cases{ \pi (n+1)  S_{CS}(A)  & if $n$ is even \cr
\pi (n+1)  S_{CS}(A)  & if $n$ is odd \cr}
\cr}
\eqno{(2.20)}
$$
where $S_{CS}(A)$ is the Chern-Simons term:
$$
S_{CS}(A)=
{1\over 8\pi^2}
\int d^3x
\epsilon_{\mu\nu\lambda}
\tr \Bigl[ 
A_\mu(x) \partial_\nu A_\lambda(x)
+  {2\over 3} A_\mu(x) A_\nu(x) A_\lambda(x)
\Bigr] .
\eqno{(2.21)}$$
For dimensional regularization, we have $n=0$, while
for a single Pauli-Villars regulator, we have $n=\pm 1$,
depending on the sign of the mass of the Pauli-Villars field.
One can obtain results with arbitrary integer $n$, if one wishes,
since one can change $n$ by two by putting two extra Pauli-Villars
fields, one Dirac boson and one Dirac fermion, with masses of opposite signs.
 
The existence of $h(A)$ cannot be deduced within perturbative calculations
such as the ones considered so far.
The lattice regularization using Wilson fermion [9], 
which gives the odd $n$ case, is manifestly gauge invariant on the lattice, 
and one can see that $h(A)$ must be there to ensure 
the gauge invariance under large gauge transformations.
One should note first that under 
a large gauge transformation, the Chern-Simons term
transforms as
$$
S_{CS}(A) \rightarrow S_{CS}(A) + \nu,
\eqno{(2.22)}
$$
where $\nu$ is the winding number of the gauge transformation.
$\exp[ik\pi S_{CS}(A)]$ is invariant under any 
large gauge transformation for even $k$, but not for odd $k$.
For odd $k$, $h(A)$ compensates the variation of the Chern-Simons term under
the large gauge transformation.
The additional term $h(A)$ is parity even and therefore it should be 0 or $\pi$.
It is 0 for perturbative gauge field, and so it cannot be seen
perturbatively.
Under a large gauge transformation with odd winding number, it changes
from 0 to $\pi$ or vice versa, thus making the total effective action 
gauge invariant.

It is known that $h(A)$ is non-local
and therefore it is natural to expect that 
the even $n$ case is connected to the odd $n$ case by 
simply adding the Chern-Simons action, which is local
in terms of the gauge field.
Then the $h(A)$ must be present in the massless case and
not in the infinite mass case for even $n$ also as in (2.20).
This implies that the total effective action for even $n$ case 
is not invariant under the large gauge transformations.
We construct a lattice formalism that corresponds to $n=0$ in Section IV
and see that the above is indeed the case.

\bigskip
\centerline{\bf III. Perturbative regularization using 
infinite number of Pauli-Villars fields}
\medskip

In the previous section, we have seen that the use of a single 
Pauli-Villars regulator gives rise to parity anomaly.
In this section we see that even within Pauli-Villars regularization,
one can make it manifestly parity invariant by using
infinite number of Pauli-Villars fields\footnote{${}^\star$}
{This issue was formerly addressed by Ref. [14].
We thank Y. Kikukawa for informing us of this work.}.
Such a formalism has been developed in even dimensions
in the context of regularizing chiral gauge theory in a gauge invariant
way [10,11].

We start with the action for a massless Dirac
fermion
$$ S  = \int d^3x \bar{\psi}(x) \dirac  \psi(x) \eqno{(3.1)}$$
and
add to it an infinite number of fermi and bose regulator
fields as follows:
$$\eqalign{
S_n^{(f)} & = \int d^3x \bar{\psi}_n(x) (\dirac+n\Lambda)  \psi_n(x) ;
\ \ \ \  n=\pm 2, \pm 4, \cdots \pm\infty \cr
S_n^{(b)} & = \int d^3x \bar{\phi}_n(x) (\dirac+n\Lambda)  \phi_n(x);
\ \ \ \ n=\pm 1, \pm 3, \cdots \pm\infty \cr}
\eqno{(3.2)}
$$
Note that this regularization is manifestly parity invariant, 
since the total action is invariant under
$$
\psi(x) \rightarrow \psi(-x);\ \ \ \ \
\bar{\psi}(x) \rightarrow -\bar{\psi}(-x);
$$
$$
\psi_n(x) \rightarrow \psi_{-n}(-x);\ \ \ \ \
\bar{\psi}_n(x) \rightarrow -\bar{\psi}_{-n}(-x);
$$
$$
\phi_n(x) \rightarrow \phi_{-n}(-x);\ \ \ \ \
\bar{\phi}_n(x) \rightarrow -\bar{\phi}_{-n}(-x);
$$
$$
A_\mu(x) \rightarrow -A_\mu(-x).
\eqno{(3.3)}
$$

The regularized vacuum polarization tensor is given by
$$ 
\Pi_{\mu\nu}^{(reg)}(p) =
{1\over 8} 
\int {d^3q \over (2\pi)^3}
\sum_{n=-\infty}^{\infty} (-1)^n
{\tr \bigl[
\gamma_\mu (-i\gamma_\lambda ({p-q\over 2})_\lambda + n \Lambda )
\gamma_\nu
(i\gamma_\rho ({p+q\over 2})_\rho +  n \Lambda ) \bigr]\over
\bigl[({p-q\over 2})^2 + (n \Lambda)^2 \bigr]
\bigl[({p+q\over 2})^2 + (n \Lambda)^2 \bigr]}.
\eqno{(3.4)}$$
The finiteness of this expression can be seen following Ref. [10,11].
Let us focus on the four-gamma term, which is linear divergent,
superficially.
Since
$$\eqalign{ &
 \sum_{n=-\infty}^\infty {(-1)^n \over
\bigl[({p-q\over 2})^2+ n^2\Lambda^2\bigr]
\bigl[({p+q\over 2})^2+ n^2\Lambda^2\bigr]
}\cr
= &{2\pi\over \Lambda (p_\mu q_\mu)}\Biggl[
{1 \over |p-q| \sinh {|p-q|\pi\over 2\Lambda}}
-{1 \over |p+q| \sinh {|p+q|\pi\over 2\Lambda}}  
\Biggr]}
\eqno{(3.5)}
$$
the divergence in the integral over the loop momentum is
sufficiently tamed by the additional contributions from the infinite number
of Pauli-Villars fields.
Since the regularization preserves manifest parity invariance,
the result should be parity invariant.
This can be checked explicitly also.
The three-gamma term which gives the parity odd term
adds up to zero, since the contributions from positive $n$ and negative $n$
cancel each other.

As can be seen above,
the situation is indeed quite similar to the one encountered
in regularizing a chiral fermion coupled to external gauge field.
In order to make the connection more explicit, 
we rewrite the above formalism as follows.
We first redefine the fields as
$$
\psi^{(+)}_{1} = \psi;
\ \ \ \  
\bar{\psi}^{(-)}_{1} = \bar{\psi};
$$
$$
\psi^{(+)}_{s+1} = {\psi_{2s} + \psi_{-2s} \over \sqrt{2}};
\ \ \ \  
\bar{\psi}^{(-)}_{s+1} = {\bar{\psi}_{2s} + \bar{\psi}_{-2s} \over \sqrt{2}};
\ \ \ \  s= 1, 2, \cdots, \infty 
$$
$$
\psi^{(-)}_{s} = {\psi_{2s} - \psi_{-2s} \over \sqrt{2}};
\ \ \ \  
\bar{\psi}^{(+)}_{s} = {\bar{\psi}_{2s} - \bar{\psi}_{-2s} \over \sqrt{2}};
\ \ \ \  s= 1, 2, \cdots,\infty 
$$
$$
\phi^{(+)}_{s} = {\phi_{2s-1} + \phi_{-(2s-1)} \over \sqrt{2}};
\ \ \ \  
\bar{\phi}^{(-)}_{s} = {\bar{\phi}_{2s-1} + \bar{\phi}_{-(2s-1)} 
\over \sqrt{2}};
\ \ \ \  s= 1,  2, \cdots ,\infty 
$$
$$
\phi^{(-)}_{s} = {\phi_{(2s-1)} - \phi_{-(2s-1)} \over \sqrt{2}};
\ \ \ \  
\bar{\phi}^{(+)}_{s} = {\bar{\phi}_{2s-1} - \bar{\phi}_{-(2s-1)} 
\over \sqrt{2}};
\ \ \ \  s= 1,  2, \cdots,\infty ,
\eqno{(3.6)}
$$
where the suffix $+$ and $-$ denote the transformation under parity.
In the following, we refer to the index $s$ as ``flavor'' index.
We define the parity doublet as
$$
\Psi_s=\pmatrix{\psi_s^{(+)}
\cr 
\psi_s^{(-)}
\cr};\ \ \ \ 
\bar{\Psi}_s = \pmatrix 
{\bar{\psi}^{(+)}_s & \bar{\psi}^{(-)}_s \cr};
\ \ \ \  s= 1,  2, \cdots,\infty ,
\eqno{(3.7)}
$$
$$
\Phi_s=\pmatrix{\phi_s^{(+)}
\cr 
\phi_s^{(-)}
\cr};\ \ \ \ 
\bar{\Phi}_s = \pmatrix 
{\bar{\phi}^{(+)}_s & \bar{\phi}^{(-)}_s \cr};
\ \ \ \  s= 1,  2, \cdots,\infty .
\eqno{(3.8)}
$$
The total action can be written as
$$
\eqalign{
S_{tot} =& \int d^3x \Bigl[
\bar{\Psi} \Gamma_\mu D_\mu \Psi
+\bar{\Phi} \Gamma_\mu D_\mu \Phi \cr \Bigr.
& + \bigl( \bar{\Psi} {\cal M}_\Psi 
{1+\Gamma_4 \over 2} \Psi + 
\bar{\Psi} {\cal M}_\Psi^\dagger {1-\Gamma_4 \over 2} \Psi \bigr)
\cr
& \Bigl. +
\bigl( \bar{\Phi} {\cal M}_\Phi {1+\Gamma_4 \over 2} \Phi 
+ \bar{\Phi} {\cal M}_\Phi^\dagger {1-\Gamma_4 \over 2} \Phi \bigr) \Bigr],
\cr}
\eqno{(3.9)}
$$
where 
$D_\mu=\partial_\mu + i A_\mu(x) $ for $\mu=1,2,3$ and 
$$\Gamma_\mu= \pmatrix{0 & \gamma_\mu \cr \gamma_\mu & 0 \cr}
~~~for~~\mu=1,2,3;\ \ \ \
\Gamma_4= \pmatrix{1  & 0 \cr 0 & - 1  \cr}.\eqno{(3.10)}$$
${\cal M}_\Psi$ and ${\cal M}_\Phi$ are infinite dimensional matrices
defined as
$$
({\cal M}_\Psi)_{st}=2s \Lambda \delta_{s+1,t}
\ \ \ \  s,t= 1,  2, \cdots,\infty ,
\eqno{(3.11)}
$$
$$
({\cal M}_\Phi)_{st}=(2s-1) \Lambda \delta_{s,t}
\ \ \ \  s,t= 1,  2, \cdots,\infty .
\eqno{(3.12)}
$$
The action is invariant under
$$
\eqalign{
& \Psi(x) \rightarrow \Gamma_4 \Psi(-x);\ \ \ \ \
\bar{\Psi}(x) \rightarrow \bar{\Psi}(-x) \Gamma_4; \cr
& \Phi(x) \rightarrow \Gamma_4 \Phi(-x);\ \ \ \ \
\bar{\Phi}(x) \rightarrow \bar{\Phi}(-x) \Gamma_4; \cr
& A_\mu(x) \rightarrow -A_\mu(-x).}
\eqno{(3.13)}
$$
In this form, the formalism is essentially the same as the one in 
Ref. [11].
Note that dim(Ker(${\cal M}_\Psi$)) =1 and
dim(Ker(${\cal M}_\Psi^\dagger$)) =0
ensures that we have a single massless Dirac fermion,
which is not paired with a parity partner.
This is possible since the flavor space is infinite dimensional.
If we had a finite number of flavors, we would have
dim(Ker(${\cal M}_\Psi$)) = dim(Ker(${\cal M}_\Psi^\dagger$)),
and the massless Dirac fermions would, therefore, have to be paired.
Note also that since the regularization does nothing to the parity odd terms,
they should be convergent by itself
in order to ensure that the regularization really works.
In even dimensions this leads to the requirement of the gauge anomaly 
cancellation [11].
In three dimensions, the parity odd term is convergent,
actually zero in the massless case.

Now, there is a large freedom 
in choosing the infinite dimensional mass matrix 
${\cal M}_\Psi$, keeping the parity invariance manifest.
As in Ref. [11], we can consider the domain-wall realization of the
mass matrix.
Let us regard the flavor index $s$ to be continuous, ranging from 
$-\infty$ to $\infty$.
Then the mass matrix can be represented as an operator acting
on a normalizable function of $s$.
Let us define it as follows.
$$
{\cal M}_{\Psi}= {\partial \over \partial s} + m_{DW}(s),
\eqno{(3.14)}
$$
$$
{\cal M}_{\Psi}^\dagger= - {\partial \over \partial s} + m_{DW}(s).
\eqno{(3.15)}
$$
$m_{DW}(s)$ has a domain wall type dependence on $s$;
{\it e.g.} $m_{DW}(s)=-m_0$ for $s<0$ and
$m_{DW}(s)=m_0$ for $s>0$.
One can check that dim(Ker(${\cal M}_\Psi$)) =1 and
dim(Ker(${\cal M}_\Psi^\dagger$)) =0 is satisfied.
The action for the fermions now reads
$$
\int d^3 x ds 
\bar{\Psi}(x,s) (\sum_{a=1}^{4}  \Gamma_a D_a 
+ m_{DW}(s) ) \Psi(x,s)
\eqno{(3.16)}
$$
where $D_4=\partial_s$.
Regarding $s$ as the fourth coordinate,
the above action describes a four-dimensional Dirac fermion
with the domain-wall mass $m_{DW}(s)$.
Thus we arrive at odd-dimensional version of the domain-wall 
approach first proposed by Kaplan [15].
Therefore we can do everything that has been done in even dimensions
in the context of chiral fermion.
Above all, a lattice formulation of a massless Dirac fermion 
with exact parity invariance must be possible using the 
overlap formalism [12], 
which we show in the next section.

\bigskip
\centerline{\bf IV. Massless fermion in the 
overlap formalism}
\medskip

In this section we formulate the overlap formalism [12] for
three-dimensional massless Dirac fermions
coupled to an external gauge field and show that it gives 
parity invariant lattice regularization.
We show that the fermion determinant is real and that only its sign
can change under gauge transformations.

Overlap formalism provides a method to define the determinant
of operators without reference to any eigenvalue problem. This
is of particular use in chiral gauge theories where the chiral
Dirac operator does not possess an eigenvalue problem. But it must be
also of use in the present situation
where reference to an eigenvalue problem seems to lead
to parity anomaly [9].

In the overlap formalism, the determinant of $\dirac$ is expressed as an
overlap of two many-body states. The two many-body states are
ground states of two many-body Hamiltonians describing non-interacting
fermions. Explicitly, the two many-body Hamiltonians are
$$\mbham^\pm =a^\dagger \ham^\pm a\eqno{(4.1)}$$
where
$$\ham^\pm = \pmatrix{ \pm m_o & \dirac \cr -\dirac & \mp m_o \cr }
\eqno{(4.2)}$$
Following [12] one can show that the
overlap of the two many-body ground states is indeed the determinant of
$\dirac$ as $m_o\rightarrow\infty$. A lattice regularization of the
overlap is obtained by a lattice regularization of the Hamiltonians
that appear in (4.2). The regularized Hamiltonians are given by
$$\ham^\pm = \pmatrix{ \wilson \pm m_o & \dirac \cr
-\dirac & -\wilson \mp m_o \cr } \eqno{(4.3)}$$
where
$$
\dirac(x\alpha i, y\beta j) =
{1\over 2}
\sum_{\mu=1}^3 \gamma_\mu^{\alpha\beta}
\Bigl [
\delta_{y,x+\hat\mu} (U_\mu(x))^{ij} -
\delta_{x,y+\hat\mu} (U^\dagger_\mu(y))^{ij}
\Bigr] \eqno{(4.4)}
$$
$$
\wilson(x\alpha i, y\beta j) =
{1\over 2} \delta_{\alpha\beta}\sum_{\mu=1}^3
\Bigl [
2 \delta_{xy}\delta_{ij} -
\delta_{y,x+\hat\mu} (U_\mu(x))^{ij} -
\delta_{x,y+\hat\mu} (U^\dagger_\mu(y))^{ij}
\Bigr] \eqno{(4.5)}
$$
and
$$\dirac^\dagger = -\dirac;\ \ \ \ \wilson^\dagger = \wilson
\eqno{(4.6)}$$
$m_o$ should be
kept fixed at a non-zero value as the continuum limit is taken.

We will define $|L\pm\wbrU$ as the ground states of  
the two Hamiltonians and $|R\pm\wbrU$ as the top states (or
ground states of $-\mbham^\pm$) of the two Hamiltonians.
The lattice formulae for the determinants are
$$
\det\dirac(U) = \wblU L-|L+\wbrU;\ \ \ \ \
\det\dirac^\dagger(U) = \wblU R-|R+\wbrU \eqno{(4.7)}$$
The formula is completely defined only if the phases of the many
body states are also specified and this is done using the
Wigner-Brillouin phase choice [12]. This choice imposes the condition
that $\wblU L\pm | L\pm \wbr1$ and $\wblU L\pm | L\pm \wbr1$ are both
real and positive for all $U$'s.

We now state and prove three Lemmas that can be combined together
to show that the overlaps in (4.7) are parity invariant and real.

\noindent{\sl Lemma 4.1.}
$$ \wblU R-|R+\wbrU = \wblU L+|L-\wbrU 
~~~~~\Bigl(\iff
\det\dirac^\dagger(U) = \{\det\dirac(U) \}^\ast \Bigr)
$$

\noindent{\sl Proof:}\hfill\break
Let
$$\Sigma = \pmatrix { 0 & I \cr I & 0 \cr };
\ \ \ \ \ \Sigma^2 =1 \eqno{(4.8)}$$
Then
$$\Sigma \ham^\pm \Sigma = - \ham^\pm\eqno{(4.9)}$$
for all gauge fields on the lattice.
Let $\psi_K^{L\pm} (x\alpha i; 1)$ be the set of eigenvectors
corresponding to all the negative eigenvalues when $U=1$.
We are free to choose the phase of these eigenvectors and we
assume that they have been chosen in such a way that
$\wbl1 L-|L+\wbr1$ is real and positive. Now we set
$$\psi_K^{R\pm} (x\alpha i; 1) = \Sigma_{\alpha\beta}
\psi_K^{L\pm} (x\beta i; 1) \eqno{(4.10)}$$
Then it is clear that $\wbl1 R-|R+\wbr1$ is also real and positive.
This completes our choice for the free many-body states that will be
used in the Wigner-Brillouin phase choice.
Let ${\cal U} (\pm;U|\pm;U^\prime)$
be the unitary matrix that relates the bases
diagonalizing $\ham^\pm(U)$ and $\ham^\pm(U^\prime)$.
Then our choice of the free states
imply that
$$\det {\cal U} (+;1|-;1) = 1 \eqno{(4.11)}$$
The Wigner-Brillouin phase choice implies that
$$\det {\cal U} (\pm;1|\pm;U) = 1 \eqno{(4.12)}$$
Since
$${\cal U} (-;U|+;U) ={\cal U} (-;U|-;1){\cal U} (-;1|+;1){\cal U} (+;1|+;U)
\eqno{(4.13)}$$
(4.11) and (4.12) imply that
$${\wblU R-|R+\wbrU \over \wblU L-|L+\wbrU }
= \det {\cal U} (-;U|+;U) =1\eqno{(4.14)}$$
and this proves the lemma.

\medskip

\noindent{\sl Lemma 4.2.}
$$ \wblU R-|R+\wbrU = \wblU L-|L+\wbrU 
~~~~~\Bigl( \iff
\det\dirac^\dagger(U) = \det\dirac(U) \Bigr)
$$

\noindent{\sl Proof:}\hfill\break
{}From (4.9) we have
$$\psi^{R\pm}_K(x\alpha i; U) = e^{i\phi^{R\pm}_K} \Sigma_{\alpha\beta}
\psi^{L\pm}_K(x\beta i; U)\eqno{(4.15)}$$
where $\psi^{R\pm_K}$ are undetermined phases. Using (4.10) we have
$$\wbl1 R\pm|R\pm\wbrU = \wbl1 L\pm|L\pm\wbrU
e^{i\sum_K \phi^{R\pm}_K} \eqno{(4.16)}$$
Since
$\wbl1 R\pm|R\pm\wbrU$ and $\wbl1 L\pm|L\pm\wbrU$ are both real
and positive, we have
$$\sum_K \phi^{R\pm}_K =0.\eqno{(4.17)}$$
Now from eqn (4.15) we have
$$\wblU R-|R+\wbrU = \wblU L-|L+\wbrU
e^{i\sum_K [\phi^{R+}_K - \phi^{R-}_K]} = \wblU L-|L+\wbrU \eqno{(4.18)}$$
where the last equality follows from eqn (4.17).
This proves the lemma.

\medskip

\noindent{\sl Lemma 4.3.}
For gauge field related by parity,
$U^\prime_\mu(x) = U^\dagger_\mu (-x-\hat\mu)$, we have
$$ \wblUp  L-|L+\wbrUp = \wblU L-|L+\wbrU 
~~~~~ \Bigl(\iff
\det\dirac(U') = \det\dirac(U) \Bigr)
$$

\noindent{\sl Proof:}\hfill\break
 Let
$$\Gamma = \pmatrix { 1 & 0 \cr 0 & -1 \cr }\eqno{(4.19)}$$
Then one has
$$\ham^\pm [x\alpha i, y \beta j; U^\prime] = \Gamma_{\alpha\gamma}
\ham^\pm [- x\gamma i, -y\delta j; U] \Gamma_{\delta\beta}
\eqno{(4.20)}$$
on the lattice.  
Therefore we have the following relation between eigenvectors.
$$\psi^{L\pm}_K(x\alpha i; U^\prime) =
e^{i\phi^{L\pm}_K} \Gamma_{\alpha\beta}
\psi^{L\pm}_K(-x\beta i; U)\eqno{(4.20)}$$
For the special case of $U=1$ the above equation becomes
$$\psi^{L\pm}_K(x\alpha i; 1) =
e^{i\chi^{L\pm}_K} \Gamma_{\alpha\beta}
\psi^{L\pm}_K(-x\beta i; 1)\eqno{(4.21)}$$
Since we have chosen $\wbl1 L-|L+\wbr1$ to be real and positive
we have
$$\sum_K[ \chi^{L+}_K - \chi^{L-}_K ] = 0\eqno{(4.22)}$$
Now from (4.20) and (4.21) we have
$$\wbl1 L\pm|L\pm\wbrUp = \wbl1 L\pm|L\pm\wbrU
e^{i\sum_K [
\phi^{L\pm}_K -
\chi^{L\pm}_K ]}\eqno{(4.23)}$$
Since
$\wbl1 L\pm|L\pm\wbrUp$ and  $\wbl1 L\pm|L\pm\wbrU $
are both real and positive it follows that
$$\sum_K [
\phi^{L\pm}_K - \chi^{L\pm}_K ]  = 0  \eqno{(4.24)}$$
{}From (4.20), we get
$$\wblUp L-|L+\wbrUp = \wblU L-|L+\wbrU
e^{i\sum_K [\phi^{L+}_K - \phi^{L-}_K ]} 
= \wblU L-|L+\wbrU
\eqno{(4.25)}$$
where the last equality follows from eqns (4.22) and (4.24).
This proves the lemma.

Lemma 1 along with Lemma 2 shows that $\wblU L-|L+\wbrU $
is real for all $U$'s on the lattice. Lemma 3 shows that it
is parity even on the lattice.

Under a gauge transformation, the Hamiltonians in (4.3) are rotated
by a unitary transformation and therefore the determinant as defined
in (4.7) can only change by a phase under a gauge transformation.
Since Lemmas 1 and 2 show that the determinant is real, it can
only change by a sign under a gauge transformation. Therefore
for small gauge transformations it must be gauge invariant. 

\figure{1}{\captionone}{5.0}

We examine if the determinant changes its sign
for some class of gauge transformations on the lattice 
numerically\footnote{${}^\star$}
{Before the revision of this work, Ref. [16] appeared, which deals with
the overlap in three dimensions in a completely independent manner. In
Ref. [16], it is clearly shown that the global anomaly is properly
reproduced by the overlap for the first time. This important point was
not realized in the first version of this paper.}.
We compute the overlap on the trivial orbit 
for a variety of gauge transformations, small, random and large, 
when the gauge group is SU(2). 
When we make the gauge transformation small by
restricting it close to unity, we find the result to be real and
positive. When we make random gauge transformations (i.e. $g_n\in$ SU(2)
is randomly chosen at each site $n$ with complete independence), we find
that the overlap is real and positive. We verified this on a $4^3$ and
$5^3$ lattice by performing 100 random gauge transformations on the
trivial orbit. 
The large gauge transformation we tried is 
$$g_n = e^{i \rho \sigma\cdot n}$$ 
with $\rho$ being a free scalar parameter, $n$ being the site on the
lattice with $n=0$ in the middle of the lattice. We then imposed
periodic boundary conditions. 
We scanned a region of $\rho$ from $0$ to $2$ and did find a region
where the overlap is negative. We plot the result in figure 1. 
The x-axis is the parameter $\rho$ and the y-axis is the phase of the
overlap relative to $g_n=1$. Note that the phase can be zero or $\pi$
since the overlap is real. One can see from the figure that for a small
window in $\rho$ the phase is $\pi$. This window is bigger when $L=5$
compared to $L=4$ indicating that the result is robust and will survive
as $L\rightarrow\infty$. 
Thus we confirmed that the overlap changes its sign under large
gauge transformations.
When one applies the formalism to a real problem keeping the parity invariance
manifest, one has to choose the fermion content so that the global gauge anomaly
vanishes.

\bigskip
\centerline{\bf V. Massive fermion in the overlap formalism}
\medskip

A mass term for the Dirac fermion can be introduced in the overlap
formalism
by modifying the Hamiltonians in (4.3) to read
$$\ham^\pm = \pmatrix{ \wilson \pm m_o & \dirac + m\cr
-\dirac + m & -\wilson \mp m_o \cr } \eqno{(5.1)}$$
where $m$ is the fermion mass. 
It is easy to see that Lemma 4.1--4.3 are now generalized as
\item{i)} $ 
\det(\dirac^\dagger(U)+m)  = \{ \det(\dirac(U) +m ) \}^\ast $
\item{ii)} $ 
\det( \dirac^\dagger(U)+m ) = \det ( \dirac(U) -m) $
\item{iii)} $ \det ( \dirac(U') + m ) = \det ( \dirac(U) - m), $

\noindent respectively.
Also one can show, following Lemma 4.5 in Ref. [12],
that
$$ \det ( \dirac(U^C) + m ) = \{ \det ( \dirac(U) + m) \}^\ast, $$
where $U_\mu^C(x)$ is the charge conjugation of the $U_\mu(x)$;
{\it i.e.,} $U_\mu^C(x) = (U_\mu(x))^\ast$.
This has an important consequence when the fermion is in 
a real ($ (U_\mu(x))^\ast = U_\mu(x) $)
or pseudo-real ($ (U_\mu(x))^\ast = R U_\mu(x) R^\dagger$, with
a unitary matrix $R$) representation,
that the fermion determinant in those cases is real even for $m \ne 0$.

Unless the fermion is in a real (or pseudo-real) representation,
the overlap formula for the massive case
leads to a complex determinant. As such
the gauge invariance will also be mildly broken but this is not
of concern for what we want to discuss here.

\figure{2}{\captiontwo}{5.0}
\figure{3}{\captionthree}{5.0}

In order to study the infinite mass limit in the overlap formalism,
we focus our attention on SU(2)
as an example. Our aim is to show that the result is given by (2.20) with
$n=0$. 
To this end, we pick a specific gauge background, namely
a constant non-abelian background given by
$$U_\mu(x) = e^{i{c\over L} \sigma_\mu}\eqno{(5.2)}$$
where $c$ is a constant. 
This problem can be essentially solved
analytically. This is because the Hamiltonians are block diagonal
in momentum space and one can therefore reduce it to a problem of
finding a many-body ground state per momentum. To compute the overlap
we still have to perform a product over momentum. On a finite lattice
this is a product over a finite set of momenta. The logarithm of the
overlap is the induced effective action. 
Computations were done with several values of $c$ and here we
illustrate the results for $c=0.01\pi$.
We compute the imaginary part of the effective action on several 
lattices with a fixed physical mass. We did this by setting
$m=m_p/L$ in equation (5.1) where $m_p$ is the fixed physical mass.
The Chern-Simons action (2.21) for the present gauge background
is equal to 
$$\alpha= {1 \over 8 \pi^2} {2 \over 3} c^3 \cdot 2 \cdot 3 ! = 10^{-6}\pi$$ 
and we expect
the imaginary part of the effective action in the $m_p\rightarrow \infty$
limit to be $\pi \alpha$ from (2.20) with $n=0$.
We plot the imaginary part of the effective action 
normalized to $\pi\alpha$ as a function
of $1/L$ for various $m_p$. As can be seen from figure 2, 
where we show the result for $m_p=6$, a straight line fits the data points well.
We use a linear fit to get the $L=\infty$ value at each $m_p$ and 
plot the extrapolated values as a function of $m_p$ in figure 3. 
As can be seen from the figure, the data approach nicely to unity as one
increases $m_p$. 

It should be noted that this result is for a gauge field
that has only zero momentum. To establish that indeed the infinite mass
limit is given by (2.20) with $n=0$ the above result has to verified
analytically for an arbitrary gauge field background. One has to be
carefull in such a computation. The result has to obtained using lattice
regularization for a fixed but finite $m_p$. Therefore $m\rightarrow 0$ as
$ L\rightarrow\infty$ in this limit. This is the case for all finite $m_p$,
One has to look at the $m_p\rightarrow\infty$ limit of the resulting
functional of gauge field and $m_p$. This limit need not correspond to
the limit $m\rightarrow\infty$ considered in [16]. This needs further
investigation.

\bigskip
\centerline{\bf VI. Summary and conclusions}
\medskip

The massless Dirac fermions in 3D can give either a parity-invariant result
or a parity-violating result depending on the regularization.
Pauli-Villars regularization as well as 
the lattice regularization using Wilson fermion is known to give the latter.
In this paper, we pointed out that regularizing massless Dirac fermion 
in odd dimensions keeping parity invariance manifest
within Pauli-Villars or lattice regularization
is essentially the same problem as regularizing chiral fermion
in even dimensions.
We have shown that the overlap formalism gives 
a parity invariant lattice formalism,
which enables a nonperturbative study of 3D gauge fermion system
in a parity-invariant setting.
The formalism can be used with any gauge groups and with
any representations for the fermion, though we have to choose them so 
that the gauge anomaly under large gauge transformation vanishes.

Among the numerous applications of our formalism,
an interesting problem which should be addressed is
whether the parity invariance of the 3D gauge-fermion system
is spontaneously broken or not,
when the gauge dynamics is switched on.
Physical implications to fractional quantum Hall effect and to 
high $T_c$ superconductivity should also be clarified.
Finally we comment that
this formalism can be extended to the case with 
massless Majorana fermion,
providing a study of three-dimensional ${\cal N}=1$ super Yang-Mills theory
on the lattice without fine-tuning [17].

\bigskip
\bigskip
\centerline{\bf Acknowledgement}
\medskip
The authors would like to thank Harald Markum and the other organizers of
the Vienna Lattice Workshop '96 which triggered the current collaboration.
R.N. would like to thank Luis Alvarez-Gaume,
Herbert Neuberger and Ann Nelson for some useful discussions. 
J.N. is grateful to Sinya Aoki, Hikaru Kawai and Yoshio Kikukawa 
for valuable comments.
The research of R.N. was supported in part by the DOE under grant
\# DE-FG03-96ER40956 and \# DE-FG06-90ER40561.

\vfill\eject

\centerline{\bf References}

\item{[1]} G.W. Semenoff and P. Sodano, Phys. Rev. Lett. 57 (1986) 1195.
\item{[2]} G. Baskaran and P.W. Anderson, Phys. Rev. B37 (1988) 580.
\item{[3]} S. Deser, R. Jackiw and S. Templeton,
Ann. Phys. (N.Y.) 140 (1982) 372.
\item{[4]} Y. Hosotani, Phys. Lett. B319 (1993) 332.
\item{[5]} A.N. Redlich, Phys. Rev. Lett. 52 (1984) 18;
\item{}
A.N. Redlich, Phys. Rev. D29 (1984) 2366;
\item{}
A.J. Niemi and G.W. Semenoff, Phys. Rev. Lett. 51 (1983) 2077.
\item{[6]} G. 't Hooft and M. Veltman, Nucl. Phys. B44 (1972) 189.
\item{[7]} P. Breitenlohner and D. Maison,
Commun. Math. Phys. 52 (1977) 11.
\item{[8]} R. Delbourgo and A.B. Waites, Phys. Lett. B300 (1993) 241;
\item{}
K. Stam, Phys. Rev. D34 (1986) 2517;
\item{}
B.M. Pimentel, A.T. Suzuki and J.L. Tomazelli,
Int. J. Mod. Phys. A7 (1992)  5307;
\item{}
T. Appelquist, M.J. Bowick, D. Karabali and L.C.R. Wijewardhana,
Phys. Rev. D33 (1986) 3774.
\item{[9]} A. Coste and M. L\"uscher, Nucl. Phys. B323 (1989) 631.
\item{[10]} S.A. Frolov and A.A. Slavnov, Phys. Lett. B309 (1993) 344.
\item{[11]} R. Narayanan and H. Neuberger, Phys. Lett. B302 (1993) 62.
\item{[12]} R. Narayanan and H. Neuberger, Nucl. Phys. B443 (1995) 305.
\item{[13]} M. Bos, Ann. Phys. (N.Y.) 181 (1988) 197;
\item{}
H. Osborn, Ann. Phys. (N.Y.) 200 (1990) 1;
\item{}
G. Giavarini, C.P Martin and F. Ruiz Ruiz,
Nucl. Phys. B381 (1992) 222;
\item{}
F. Ruiz Ruiz and P. van Nieuwenhuizen, Nucl. Phys. B486 (1997)
443.
\item{[14]} T. Kimura, Prog. Theor. Phys. 92 (1994) 693.
\item{[15]} D.B. Kaplan, Phys. Lett. B288 (1992) 342.
\item{[16]} Y. Kikukawa and H. Neuberger, 
{\it Overlap in Odd Dimensions}, hep-lat/9707016.
\item{[17]} N. Maru and J. Nishimura, 
{\it Lattice Formulation of Supersymmetric Yang-Mills 
Theories without Fine-Tuning}, hep-th/9705152.

\vfill\eject



\end